\documentclass[prl,twocolumn,epsf,psfig]{revtex4}
\usepackage{graphicx}
\DeclareGraphicsExtensions{.pdf,.png,.gif,.jpg}
\usepackage{float}
\usepackage{subfig}
\usepackage{epsfig}
\usepackage{wrapfig}

\begin{document}

\title{An Effective Series Expansion to the Equation of State of Unitary Fermi Gases}

\author{Theja N. De Silva}
\affiliation{Department of Chemistry and Physics,
Augusta University, Augusta, GA 30912, USA.}
\begin{abstract}
Using universal properties and a basic statistical mechanical approach, we propose a general equation of state for unitary Fermi gases. The universal equation of state is written as a series solution to a self consistent integral equation where the general solution is a linear combination of Fermi functions. First, by truncating our series solution to four terms with already known exact theoretical inputs at limiting cases, namely the first \emph{three} virial coefficients and using the Bertsch parameter as a free parameter, we find a good agreement with experimental measurements in the entire temperature region in the normal state. This analytical equation of state agrees with experimental data up to the fugacity $z = 18$, which is a vast improvement over the other analytical equations of state available where the agreements is \emph{only} up to $z \approx 7$. Second, by truncating our series solution to four terms again using first \emph{four} virial coefficients, we find the Bertsch parameter $\xi =0.35$, which is in good agreement with the direct experimental measurement of $\xi =0.37$. This second form of equation of state shows a good agreement with self-consistent T-matrix calculations in the normal phase.
\end{abstract}

\maketitle

\section{I. Introduction}

Strongly interacting Fermi particle systems are common throughout nature. Examples include quark-gluon plasmas in the early universe, nuclear matter in neutron stars, and condensed matter electronic compounds. Due to the flexibility of cold-atom experiments and possibility of tuning the interaction from a weakly attractive regime to a weakly repulsive regime through an infinitely strong interacting regime, the thermodynamics of ultra-cold Fermi atoms has been in the center of experimental investigations. In cold atom experiments, the inter-atomic interaction between neutral Fermi atoms can be controlled using the Feshbach resonance~\cite{fbr1, fbr2, fbr3, fbr4}. This is done by adjusting the two-body s-wave scattering length between two fermions in different hyperfine states. At low densities and ultra-cold temperatures, only isotropic and short-range s-wave scattering between particles can take place. Therefore scattering can be solely characterized by the s-wave scattering length $a_s$. The system is called unitary when the scattering length is adjusted to be infinitely large~\cite{treview}. For unitary fermions with zero range interacting systems, such as ultra-cold neutral atoms, the inter particle distance sets the only length scale. As a result, the details of the inter atomic interaction are not important when it’s come to the physical properties. At this unitary limit, the system is expected to show universal behavior in both static and dynamic properties, regardless of the specific system~\cite{ho}.

Recently, there have been exciting experimental efforts in realizing ultra-cold unitary Fermi systems to study the behavior of universal fermions~\cite{ereview}. As a result of the unprecedented controllability of ultra-cold atomic systems, several experimental groups have achieved the strongly interacting regime of fermions by tuning an external magnetic field across a collisional Feshbach resonance. By doing so, they have tuned the two-body s-wave scattering length $a_s$ from small positive values to small negative values through positive and negative infinities. This tuning leads them to observe the crossover from Bardeen-Cooper-Schieffer (BCS) superfluids to Bose-Einstein condensates (BEC) of two-component fermions at low temperatures. The strongly interacting regime or the unitary regime is denoted by the condition $k_F|a_s| >> 1$, where $k_F = (3 \pi^2 n)^{1/3}$ is the Fermi wave-vector with $n$ being the fermion density. At the Feshbach resonance or unitarity the s-wave scattering length is infinity, consequently the Fermi system is in the unitary regime as the dimensionless interaction parameter $k_F |a_s|$ is very large. In contrast, the nuclear matter in neutron starts is also in the unitary regime with the condition $k_F|a_s| >> 1$, because of the fact that $k_F$ is very large due to the large density of nuclear matter. Perhaps strongly correlated electronic matter, such as high-temperature superconducting compounds are also in the universal regime as the interaction between electrons in these materials are very strong. The experimental front of studying the strongly interacting fermions first started with observation of the stability of trapped fermions~\cite{ohara}. The collective excitations were later measured across the BCS-BEC crossover region~\cite{kina, bart}. After the superfluidity in the BCS-BEC crossover region has been observed~\cite{sf1, sf2}, various universal properties have been studied in the strongly interacting regime~\cite{up1, up2, up3}. Since then a remarkable experimental progress has been archived in the field of ultra-cold Fermi gases~\cite{history}. The universal thermodynamics and the equation of state at the unitary limit have been in the focus of recent experimental investigation as those provide a broader understanding of the universal physics associates with strong interactions~\cite{Ekasi7, Ekasi9, Ekasi9b}.

In theoretical point of view, the strongly interacting fermions are challenging. Due to the absence of a small parameter, the perturbative methods are inapplicable. Significant fluctuations at strongly interacting fermions make mean-field approaches also inappropriate. However, there has been several attempts to understand the strongly interacting behavior using numerical, phenomenological, and mean-field approaches. These include strong coupling theories~\cite{ohashi, engel, pera, chen, hu2, liu, hauss, diener, combe, gubbe}, Monte Carlo methods~\cite{astra, bulgac, akkine, buro, carl, houck}, modified mean-field attempts~\cite{mfe}, and some phenomenological theories~\cite{pht1, pht2, pht3}.

For zero-range interacting fermions, Tan derived a set of exact universal relations~\cite{tan1, tan2, tan3}. Tan's dramatic universal relations do not depend on the details of the interaction potential. They are applicable for broad situations: homogeneous or trapped systems, many-body or few-body systems, superfluid or normal phases, and finite or zero temperatures. These exact universal Tan relations connect microscopic properties of the zero-range strongly interacting Fermi systems to thermodynamic quantities. The connection is made through a single quantity termed contact which is a measure of the density of fermion pairs at larger momentum. Tan relations were later re-derived by using a renormalization scheme in the quantum field theoretical framework~\cite{braaten1, braaten2}, by using a lattice model to regularize the singularity~\cite{werner1, werner2}, by using a nonlocal quantum field theory~\cite{zhang}, and by using a Schrodinger formalism~\cite{combe}. These Tan relations were later experimentally verified by two experimental groups~\cite{g1, g2}.

The quantum virial cluster expansion for unitary Fermi gases has been a valuable approach to study high temperature limit of the system. In virial cluster expansion, the equation of state of the system is expanded in powers of the fugacity $z = e^{\beta \mu}$, where $\mu$ is the chemical potential and $\beta = 1/k_BT$ is the dimensionless inverse temperature. Here $k_B$ is the Boltzmann constant. The equation of state of a two-component Fermi system, the pressure in this case is written in the form,

\begin{eqnarray}
P(T, \mu) = \frac{2}{\beta \lambda^3} \sum_{n =1}^{\infty} b_n z^n,
\end{eqnarray}

\noindent where $\lambda = \sqrt{2 \pi \beta \hbar^2/m}$ is the thermal de Broglie wavelength with a fermion mass $m$ and the Plank constant $h = 2 \pi \hbar$. As a result of the universality, the virial coefficients $b_n$'s are temperature independent at unitarity. So far \emph{only} the first four virial coefficients, $b_1$, $b_2$, $b_3$, and $b_4$ have been calculated~\cite{viri1, viri2, newviri, newb4}.

In this paper, we combine one of the Tan's relations (which becomes an universal relation at the unitary limit) with basic statistical mechanics to derive a self-consistent integral equation for the equation of state for unitary Fermi gases. A general solution to this integral equation is written as a linear combination of Fermi functions. By truncating our series solution to four terms with inputs from exact limiting cases, we discuss two different versions of our equation of state. First, using first three virial coefficients and using the Bertsch parameter as a free fitting parameter, we find a good good agreements with experimentally obtained equation of state in the entire normal state. Second, using first four virial coefficients, we calculate the Bertsch parameter and find a very good agreement with the direct experimental measurements at Massachusetts Institute of Technology. Further, this second version of our equation of state agrees with the equation of state obtained by the self-consistent T-matrix calculations in the normal state~\cite{tmat}. The paper is organized as follows. In section II, we discuss our statistical mechanics approach and provide the detail derivation of the self consistent integral equation for the equation of state. In section III, we provide the solution to this integral equation and compare it with recent experimental measurements and self-consistent T-matrix calculations. Finally in section IV, we summarize our results with a discussion.

\section{II. Formalism and the Equation of State}

From a statistical point of view, the macroscopic thermodynamic properties of the system is fully captured in the partition function. The finite temperature thermodynamic potential $\Omega$ at temperature $T$ and volume $V$ is related to the partition function through $\Omega = -k_BT \ln Z_G$, where the grand canonical partition function $Z_G = Tr[e^{-(\hat{H}-\mu \hat{N}) \beta}]$ with $\hat{H}$ and $\hat{N}$ being the interacting Hamiltonian and particle number operators, respectively. For a spatially homogeneous system, the pressure $P$, is related to the thermodynamic potential as $\Omega = -PV$. This can be casted in terms of the partition function as $\beta PV = \ln Z_G$. Taking the derivative with respect to the inverse temperature $\beta$, we have

\begin{eqnarray}
\frac{\partial}{\partial \beta}(\beta PV) = -\langle \hat{H} - \mu \hat{N} \rangle,
\end{eqnarray}

\noindent where $\langle \hat{X} \rangle = Tr[ \hat{X} e^{-(\hat{H}-\mu \hat{N})\beta}]/Z_G$ is the expectation value of the operator $\hat{X}$. Meantime, one of the Tan's relations that connects the energy $\langle \hat{H} \rangle = E$, the pressure $P$, and the contact $c$,

\begin{eqnarray}
PV = \frac{2}{3}E + \frac{\hbar^2}{12 \pi m a_s}cV.
\end{eqnarray}

\noindent For unitary fermions $a_s \rightarrow \infty$, this reduces to the well-known universal equation of state $PV =2/3 E$. Replacing the right hand side of Eq. (2) with this Tan's relation and then distributing the differentiation, we have
\begin{eqnarray}
\beta \frac{\partial P}{\partial \beta} = -\frac{5}{2}P + \mu n + \frac{\hbar^2}{8 \pi m a_s}c,
\end{eqnarray}

\noindent where $n = \langle \hat{N} \rangle/V$ is the particle number density. For non-interacting fermions the contact $c$ does not exist and for unitary fermions $a_s \rightarrow \infty$. Thus, the last term in Eq. (4) vanishes at both of these limits. By dropping the last term, changing the variable $\beta$ to temperature $T$, and separating the variables, we derive a self consistent integral equation for the finite temperature pressure $P(T, \mu)$ of a homogeneous Fermi particle system,

\begin{eqnarray}
P(T, \mu) = P(0,\mu) +\frac{5}{2}\int_0^T\frac{P(x, \mu)}{x}dx \nonumber \\
-\mu \frac{\partial}{\partial \mu} \int_0^T \frac{P(x, \mu)}{x}dx.
\end{eqnarray}

\noindent Here we have replace the number density $n$ by $\partial P/\partial \mu$. So far no any approximations were made and this self consistent integral equation is valid for both non-interacting fermions and unitary fermions with zero-range interacting potentials. Even though, the contact term has disappeared in eq. (5), the interaction effects of the unitary fermions are effectively included inside the pressure $P$. Nevertheless, the temperature dependence of contact parameter at unitarity has been calculated and measured~\cite{ucontact1, ucontact2, ucontact3, ucontact4}. Notice that the zero temperature equation of state or the pressure $P(0, \mu)$ has already been known from various analytical and numerical calculations. At unitarity, the zero temperature pressure is written in the form $P(0, \mu) = 8 \alpha /(15 \sqrt{\pi}) \xi^{-3/2} \mu^{5/2}$, where $\alpha = [m/(\hbar^2 \pi 2^{1/3})]^{3/2}$ and the Bertsch parameter is defined as the ratio of ground state energies of unitary fermions and non-interacting fermions, $\xi = E_u/E_{free}$. Various theoretical calculations shows that the value of $\xi$ ranges from 0.2 to 0.6, with most predictions in the range $0.3-0.4$~\cite{bour, drut}. The earliest calculation from fixed-node diffusion Monte Carlo (MC) shows $\xi = 0.44$ for smaller systems~\cite{kasi1} and $\xi = 0.42$ for larger systems~\cite{kasi2, kasi3}. Later, several fixed-node MC calculations suggest upper bounds for $\xi$ at 0.43~\cite{kasi4}, 0.38~\cite{kasi5}, and 0.21~\cite{bour}. Meantime, a restricted path-integral MC~\cite{kasi6} and a sign-restricted mean-field lattice calculations~\cite{kasi7} yield $\xi = 0.49$. In addition to these numerical approaches, several analytical techniques, such as mean-field theories~\cite{kasi8}, variational approaches~\cite{kasi9}, saddle point methods~\cite{kasi10}, density functional theories~\cite{kasi11}, and renormalization group flow methods~\cite{kasi12} have been used to calculate the Bertsch parameter. Further, various series expansion methods have been used to calculate $\xi$~\cite{kasi13, kasi14, kasi15, kasi16, kasi17, kasi18, kasi19, kasi20, kasi21, kasi22, kasi23}. Recent cold-atom experiments have attempted to measure the value of $\xi$ using various measurement techniques and found the value of $\xi$ ranges between 0.32 and 0.51~\cite{Ekasi7, Ekasi9, Ekasi1, Ekasi2, Ekasi3, Ekasi4, Ekasi5, Ekasi6, Ekasi8}. Currently, the accepted value of $\xi = 0.37$ is from the most recent direct experimental measurements at Massachusetts Institute of Technology (MIT) experiment~\cite{Ekasi7}. This value is well supported by the improved fixed-node Monte-Carlo calculations~\cite{kasi5, carlsonNew}. As we will show below, our theory crucially depends on this Bertsch parameter $\xi$.

By investigating our self consistent integral equation in Eq. (5), we find a general solution in the form of a linear combination of Fermi functions,

\begin{eqnarray}
P(T, \mu) = \sum_{\nu \leq 5/2} A_{\nu} (k_BT)^{5/2} f_\nu(z),
\end{eqnarray}

\noindent where $f_\nu(z)$ is the well known Fermi function defined as
\begin{eqnarray}
f_\nu(z) =\frac{1}{\Gamma(\nu)}\int^\infty_0\frac{x^{\nu-1}}{z^{-1}e^x + 1}dx,
\end{eqnarray}

\noindent with Gamma function $\Gamma(\nu)$, and $A_\nu$'s are unknown constants at the moment. Notice that the half integer values of $\nu$ in $A_\nu$'s are restricted to be equal or below $5/2$, otherwise the solution diverges as $T \rightarrow 0$. For a non-interacting ideal Fermi system, $A_\nu = 0$ for all $\nu$ except $\nu =5/2$, where $A_{5/2} = \alpha$. This general equation of state for unitary Fermi gases is the main result in this paper. In literature, different versions of this solution has been used as phenomenological equations of state for unitary fermions~\cite{pht1, pht2, pht3}. In the following section, we restrict our series solutions to few terms and evaluate $A_\nu$'s based on already available theoretical findings at limiting cases, and then compare our theory with the most recent experimental measurements and self-consistent T-matrix theory.

By investigating our integral equation (5), we find that the pressure given in the virial expansion in eq. (1) is also a solution of eq. (5). Therefore the proposed solution in Eq. (6) can be considered as an another effective series solution to the equation of state.

\section{III. Results and comparisons}

 \begin{figure*}
\includegraphics[width=\textwidth]{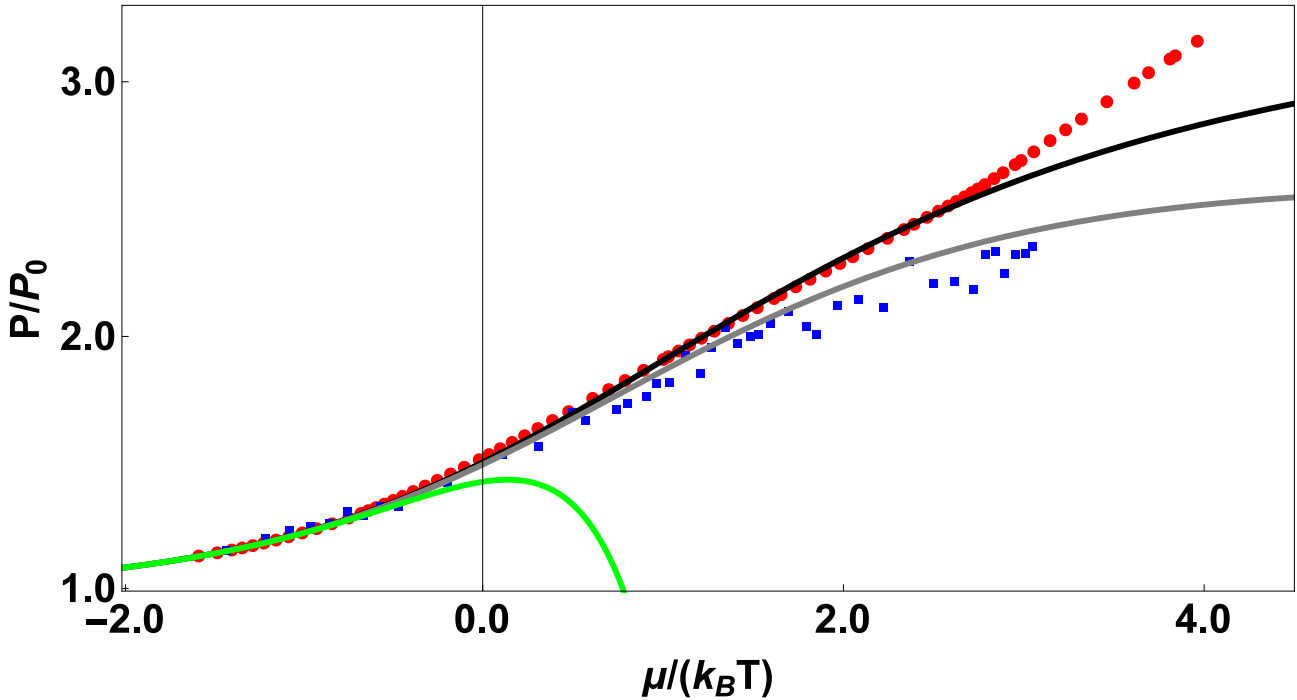}
\caption{(color online)Pressure of a unitary Fermi gas as a function of $\mu/k_BT$, normalized by the pressure of a non-interacting Fermi gas at the same chemical potential $\mu$ and temperature $T$. The red dots and blue squares are the experimental measurements at MIT and ENS, respectively. The green solid line is the third order virial cluster expansion results. The black and Gray solid lines are the results from our theoretical equation of state for selected values of Bertsch parameters.}\label{EOS}
\end{figure*}

 \begin{figure*}
\includegraphics[width=\textwidth]{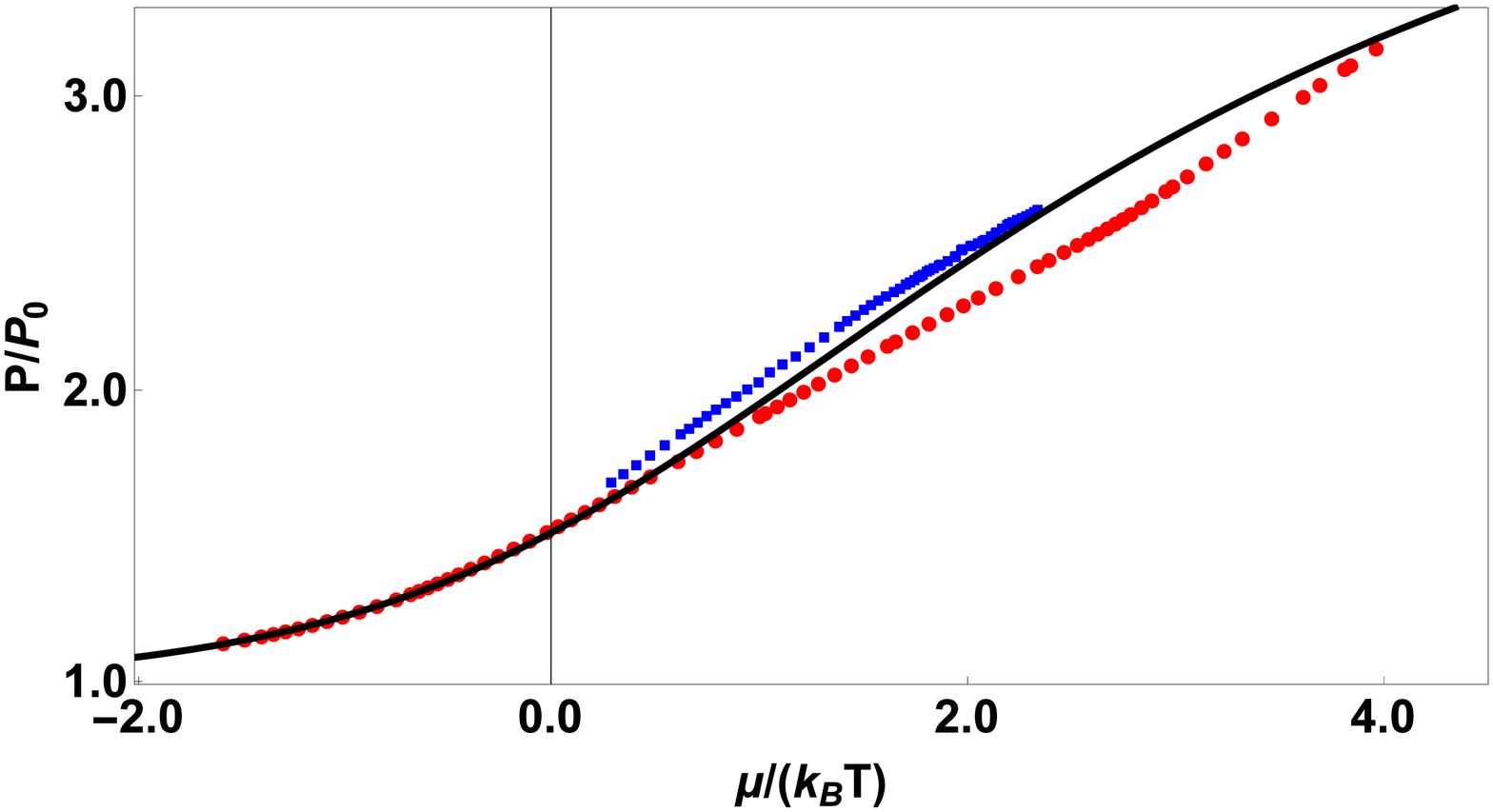}
\caption{(color online)Pressure of a unitary Fermi gas as a function of $\mu/k_BT$, normalized by the pressure of a non-interacting Fermi gas at the same chemical potential $\mu$ and temperature $T$. The red dots are the experimental measurements at MIT~\cite{Ekasi7}. The black solid line is our theoretical equation of state without any free parameters, but we truncate our series solution to four terms using first four virial coefficients. The blue squares are the normal state self-consistent T-matrix calculation~\cite{tmat}. The Bertsch parameter extracted from the first term in our solution ($A_{5/2}$) is $\xi = 0.35$.}\label{EOS2}
\end{figure*}

The zero temperature properties are precisely determined by the first term with the Fermi function $f_{5/2}(z)$. Taking the zero temperature limit in our solution, we find $A_{5/2} = \alpha \xi^{-3/2}$. Thus, our general series solution provides the exact zero temperature properties of the unitary Fermi gasses, provided the fact that $\xi$ is accurately known. By using the high temperature series expansion to our equation of state and then comparing it with the quantum virial cluster expansion, we find the $n$'th order virial coefficient $b_n$ is in the form,

\begin{eqnarray}
b_n = (-1)^{n+1} \sum_{\nu \leq 5/2} n^{-\nu} A_{\nu}.
\end{eqnarray}

\noindent As we mentioned before, the first four virial coefficients have already been calculated for homogeneous unitary Fermi gases. The latest virial cluster expansion predicts $b_1 = 1$, $b_2 = 3 \sqrt{2}/8$, and $b_3 = -0.29095295$~\cite{viri1, viri2}. Though most recent theoretical calculations of the fourth virial coefficient is in reasonable agreement with each other and with MIT experimental value ~\cite{newviri, newb4, Ekasi7}, none of the earlier calculations agree with each other~\cite{b4w1, b4w2}. In this paper, we use one of the most recent theoretical prediction of $b_4 = 0.0307$~\cite{newb4}. \emph{First}, we truncate our general series solution to four terms,

\begin{eqnarray}
P(T, \mu) = \sum_{\nu = 5/2, 3/2, 1/2, -1/2} A_{\nu} (k_BT)^{5/2} f_\nu(z),
\end{eqnarray}

\noindent where we have considered only half integer values of $\nu$ as we are dealing with fermions. Using \emph{only} the first three virial coefficients and the value of $A_{5/2}$, we find $A_{3/2} = [3b_1 + 6\sqrt{2} b_2 + 3 \sqrt{3} b_3 -11 \xi^{-3/2}/6] \alpha$, $A_{1/2} = [-5b_1/2 -8\sqrt{2} b_2 -9\sqrt{3}b_3/2 + \xi^{-3/2}] \alpha$, and $A_{-1/2} = [(3b_1 + 12\sqrt{2}b_2 + 9\sqrt{3} b_3 - \xi^{-3/2})] \alpha/6$. Note that these four coefficients are sensitive to the Bertsch parameter $\xi$ as we have already used the exact zero temperature limit. In this four term series solution, we keep $\xi$ as a free parameter to compare with the experimental results. The resulting equation of state from Eq. (9) is shown in FIG~.\ref{EOS} with recent experimental data. The FIG~.\ref{EOS} shows the pressure of a unitary Fermi system normalized by its non-interacting counterpart at the same chemical potential $\mu$ and temperature $T$, $P_0 = \alpha (k_BT)^{5/2} f_{5/2}(z)$. The red dots are the Massachusetts Institute of Technology (MIT) experimental measurements by \textbf{Ku \emph{et al}}~\cite{Ekasi7}. The blue squares are the Ecole Normale Sup\'{e}rieure (ENS) experimental measurements by \textbf{Nascimb\`{e}ne \emph{et al}}~\cite{Ekasi9}. The green solid line is the third order virial cluster expansion results. The black and gray solid lines are the results from our theoretical equation of state for two different values of Bertsch parameter, $\xi = 0.44$~\cite{kasi3, kasi4, kasi7} and $\xi = 0.59$~\cite{kasi18}, respectively. These two different values of the Bertsch parameter were chosen from the theoretically predicted range to best suited the experimental measurements. The input parameters in our theory so far are theoretically known first three virial coefficients and the Bertsch parameter, yet our theory shows a very good improvement over the third order virial cluster expansion. As can be seen from FIG.~\ref{EOS}, with use of a proper Bertsch parameter, our theory reasonably agrees with the experimental data all the way down to low temperatures where the fugacity $z \approx 18$. This is in contrast with the third order virial cluster expansion method where the virial cluster expansion theory and experimental measurements agree \emph{only} up to $z \approx 0.7$. All other proposed phenomenological equations of state show agreement with experimental data \emph{only} up to $z \approx 7$~\cite{pht1,pht2, pht3}. We tried the fourth virial coefficient to include the fifth term in our series solution, however we find that the agreement is not as good as with four term solution. Hence, in the following we use the fourth virial coefficient to find the equation of state assuming the Bertsch parameter is not known and find it through our resultant equation of state.

\emph{Next}, we again restrict our series solutions to the first four terms, $\nu = 5/2, 3/2$, $1/2$, and $-1/2$ and calculate these coefficients using Eq. (8) with reliably known first \emph{four} virial coefficients, without using the Bertsch parameter. We will calculate the Bertsch parameter from the resultant coefficient $A_{5/2}$. We find series coefficients for this case,  $A_{5/2} = 4.842$, $A_{3/2} = -2.889$, $A_{1/2} = -1.390$, and $A_{-1/2} = 0.437$. The resulting equation of state is shown in FIG~.\ref{EOS2} with recent MIT experimental data~\cite{Ekasi7} and normal state self-consistent T-matrix theory~\cite{tmat}. Notice that we have not used any free parameters except already known first \emph{four} virial coefficients to produce this result. Even though the theoretical results slightly deviate from the experimental data at intermediate temperatures, the theoretical results converge to the experimental data at a lower temperature. Using the value of $A_{5/2} = \alpha \xi^{-3/2} $, we find the Bertsch parameter $\xi = 0.35$. This is in remarkable agreement with the most recent direct experimental value of $\xi = 0.37$ at MIT experiment~\cite{Ekasi7} and recent theoretical upper bounds~\cite{nkasi1, nkasi2,nkasi3,nkasi4}. In addition, we compare our equation of state with the self-consistent T-matrix theory~\cite{tmat} and we find a good good agreement as shown in FIG~.\ref{EOS2}.

\section{IV. DISCUSSION AND SUMMARY}

Cold-atom experiments provide valuable insights into our understanding of strongly interacting matter and they provide benchmarks for the theoretical description of such matter. The experimental equation of state of unitary fermions have been measured at MIT and ENS, however the low temperatures experimental data is not in agreement with each other. Meantime, MIT experiment directly measures the Bertsch parameter $\xi$ and shows a good agreement with most recent theoretical calculations. The theoretical calculation of $\xi$ is challenging due to the fact that it is an intrinsically non-perturbative problem. Various theoretical predictions give a wide range for the value of $\xi$.

In addition to the non-interacting fermions, the proposed equation of state is valid only at unitarity, where the range of interaction is zero and the s-wave scattering length is infinite. Unfortunately, our solution involves infinite number of terms  which seems crucially depend on the Bertsch parameter $\xi$. Nevertheless, one can truncate our series solution at a finite order by theoretical or experimental inputs at limiting cases.

In summary, we have derived an approximate general equation of state for unitary Fermi gases. This is done by constructing a self consistent integral equation for the equation of state of universal fermions by combining a universal relation and basic statistical mechanics. The derived universal equation of state is a linear combination of Fermi functions. Then truncating our series solution to four terms using reliably known first three virial coefficients and using the Bertsch parameter $\xi$ as a fitting parameter, we find a good agreement with experimental measurements up to the fugacity $z \approx 18$. This is a vast improvement over the proposed phenomenological approaches where those approaches deviate from experimental data around $z \approx 7$. The importance of our four-term analytical equation of state is that all the thermodynamics properties can be accurately derived without heavy numerical calculations up to the fugacity $z \approx 18$, where the temperature is already very low. Further, using the first four virial coefficients in our equation of state of unitary fermions without any fitting parameters, we extract the Bertsch parameter and find $\xi = 0.35$. The resulting equation of state shows a good agreement with normal state self-consistent T-matrix theory. We anticipate that the our equation of state can be extended to validate for the superfluid phase by including more terms in the general solution.

\section{V. Acknowledgments}

We are grateful to Sylvain Nascimbene, Mark Ku, and their collaborators for sharing their experimental data with us. We also thank Wilhelm Zwerger, Bernhard Frank and, their collaborators for sending and sharing their self-consistent T-matrix calculations. We further acknowledge valuable communications with Martin Zwierlein.


\begin{references}
\bibitem{fbr1}  M. Houbiers, H. T. C. Stoof, W. I. McAlexander, and R. G. Hulet, Phys. Rev. A \textbf{57}, R1497 (1998).
\bibitem{fbr2} K. Dieckmann, C. A. Stan, S. Gupta, Z. Hadzibabic, C. H. Schunck, and W. Ketterle, Phys. Rev. Lett. \textbf{ 89}, 203201 (2002).
\bibitem{fbr3} S. Jochim, M. Bartenstein, A. Altmeyer, G. Hendl, C. Chin, J. Hecker Denschlag, and R. Grimm, Phys. Rev. Lett. \textbf{91}, 240402 (2003).
\bibitem{fbr4} T. Bourdel, J. Cubizolles, L. Khaykovich, K. M. F. Magalhaes, S. J. J. M. F. Kokkelmans, G. V. Shlyapnikov, and C. Salomon, Phys. Rev. Lett. \textbf{91}, 020402 (2003).
\bibitem{treview} For example, see theoretical review articles by, E. Braaten, Universal relations for fermions with large scattering lengths The BCS–BEC Crossover and the Unitary Fermi Gas edited, W. Zwerger (Heidelberg: Springer) pp 193–231 (2012); M. Randeria and E. Taylor, Annu. Rev. Condensed Matter Phys. 5 209–32 (2014).
\bibitem{ho} T.-L. Ho, Phys. Rev. Lett. \textbf{92}, 090402 (2004).
\bibitem{ereview} For example, see experimental review article by Ketterle W and Zwierlein M, Making, probing and understanding ultracold Fermi gases Proc. Int. School of Phys. ‘Enrico Fermi’, Course CLXIV (Varenna, Italy, 20–30 June 2006) ed M Inguscio et al (Amsterdam: IOS Press) pp 247–422 (2008).
\bibitem{ohara}  K. M. O’Hara, S. L. Hemmer, M. E. Gehm, S. R. Granade, and J. E. Thomas, Science \textbf{298}, 2179 (2002).
\bibitem{kina} J. Kinast, S. L. Hemmer, M. E. Gehm, A. Turlapov, and J. E. Thomas, Phys. Rev. Lett. \textbf{92}, 150402 (2004).
\bibitem{bart} M. Bartenstein, A. Altmeyer, S. Riedl, S. Jochim, C. Chin, J. Hecker Denschlag, and R. Grimm, Phys. Rev. Lett. \textbf{92}, 203201 (2004).
\bibitem{sf1} M.W. Zwierlein, J.R. Abo-Shaeer, A. Schirotzek, C.H. Schunck, W. Ketterle, Nature \textbf{435}, 1047 (2005).
\bibitem{sf2} C. A. Regal and D. S. Jin, Phys. Rev. Lett. \textbf{90}, 230404 (2003).
\bibitem{up1}  J. Kinast, A. Turlapov, J.E. Thomas, Q.J. Chen, J. Stajic, K. Levin, Science \textbf{307}, 1296 (2005).
\bibitem{up2} L. Luo, B. Clancy, J. Joseph, J. Kinast, J.E. Thomas, Phys. Rev. Lett. \textbf{98}, 080402 (2007).
\bibitem{up3} C. Cao, E. Elliott, J. Joseph, H. Wu, J. Petricka, T. Schäfer, J.E. Thomas, Science \textbf{331}, 58 (2011).
\bibitem{history} For example, see book chapter and reference there in by, M. Zwierlein, Novel Superfluids, volume 2 Edited by K. Bennemann and J. B. Ketterson ( Oxford) Chapter 18 (2015).
\bibitem{Ekasi7} Mark J. H. Ku, Ariel T. Sommer, Lawrence W. Cheuk, and Martin W. Zwierlein, Science \textbf{335}, 563 (2012).
\bibitem{Ekasi9} S. Nascimbene, N. Navon1, K. J. Jiang1, F. Chevy and C. Salomon, Nature \textbf{463}, 1057 (2010).
\bibitem{Ekasi9b} M. Horikoshi, S. Nakajima, M. Ueda, T. Mukaiyama, Science \textbf{327}, 442 (2010).
\bibitem{ohashi} Y. Ohashi and A. Griffin, Phys. Rev. Lett. \textbf{89}, 130402 (2002); Phys. Rev. A \textbf{67}, 063612 (2003)
\bibitem{engel} J. R. Engelbrecht, M. Randeria, and C. A. R. Sá de Melo, Phys. Rev. B \textbf{55}, 15153 (1997).
\bibitem{pera} A. Perali, P. Pieri, L. Pisani, and G. C. Strinati, Phys. Rev. Lett. \textbf{92}, 220404 (2004).
\bibitem{chen} Q. J. Chen, J. Stajic, S. Tan, and K. Levin, Phys. Rep. \textbf{412}, 1 (2005).
\bibitem{hu2} H. Hu, X.-J. Liu, and P. D. Drummond, Europhys. Lett. \textbf{74}, 574 (2006).
\bibitem{liu} X.-J. Liu and H. Hu, Europhys. Lett. \textbf{75}, 364 (2006).
\bibitem{hauss} R. Haussmann, W. Rantner, S. Cerrito, and W. Zwerger, Phys. Rev. A \textbf{75}, 023610 (2007).
\bibitem{diener} R. B. Diener, R. Sensarma, and M. Randeria, Phys. Rev. A \textbf{77}, 023626 (2008).
\bibitem{combe} R. Combescot, F. Alzetto, and X. Leyronas, Phys. Rev. A \textbf{79}, 053640 (2009).
\bibitem{gubbe} K. B. Gubbels and H. T. C. Stoof, Phys. Rev. A \textbf{84}, 013610 (2011).
\bibitem{astra} G. E. Astrakharchik, J. Boronat, J. Casulleras, S. Giorgini, Phys. Rev. Lett. \textbf{93}, 200404 (2004).
\bibitem{bulgac} A. Bulgac, J. E. Drut, and P. Magierski, Phys. Rev. Lett. \textbf{96}, 090404 (2006).
\bibitem{akkine} V. K. Akkineni, D. M. Ceperley, and N. Trivedi, Phys. Rev. B 76, 165116 (2007).
\bibitem{buro} E. Burovski, E. Kozik, N. Prokof’ev, B. Svistunov, and M. Troyer, Phys. Rev. Lett. \textbf{101}, 090402 (2008).
\bibitem{carl} J. Carlson and S. Reddy, Phys. Rev. Lett. \textbf{100}, 150403 (2008).
\bibitem{houck} K. Van Houcke, F. Werner, E. Kozik, N. Prokof’ev, B. Swistunov, M. Ku, A. Sommer, L. W. Cheuk, A. Schirotzek, and M. W. Zwierlein, Nature Phys. \textbf{8}, 366 (2012).
\bibitem{mfe} Erik M. Weiler and Theja N. De Silva, Phys. Rev. A 87, 013602 (2013).
\bibitem{pht1} R. K. Bhaduri, W. van Dijk, and M. V. N. Murthy, Phys. Rev. Lett. \textbf{108}, 260402 (2012).
\bibitem{pht2} M.V.N. Murthy, M.Brack, and R.K.Bhaduri, Pramana \textbf{82}, 985 (2014).
\bibitem{pht3} R. K. Bhaduri, W. van Dijk, and M. V. N. Murthy, Phys. Rev. A \textbf{88}, 045602 (2013).
\bibitem{tan1} Shina Tan, Ann. Phys. \textbf{323}, 2952 (2008).
\bibitem{tan2} Shina Tan, Ann. Phys. \textbf{323}, 2971 (2008).
\bibitem{tan3} Shina Tan, Ann. Phys. \textbf{323}, 2987 (2008).
\bibitem{braaten1} E. Braaten and L. Platter, Phys. Rev. Lett. \textbf{100}, 205301 (2008).
\bibitem{braaten2} E. Braaten and L. Platter, Phys. Rev. A 78, 053606 (2008).
\bibitem{werner1} F. Werner, Phys. Rev. A \textbf{78}, 025601 (2008).
\bibitem{werner2} F. Werner, L. Tarruell, Y. Castin, Eur. Phys. J. B \textbf{68}, 401 (2009).
\bibitem{zhang} S. Zhang and A. J. Leggett, Phys. Rev. A \textbf{79}, 023601 (2009).
\bibitem{g1} J. T. Stewart, J. P. Gaebler, T. E. Drake, and D. S. Jin, Phys. Rev. Lett. \textbf{104}, 235301 (2010).
\bibitem{g2} E. D. Kuhnle, H. Hu, X.-J. Liu, P. Dyke, M. Mark, P. D. Drummond, P. Hannaford, C. J. Vale, Phys. Rev. Lett. \textbf{105}, 070402 (2002).
\bibitem{viri1} Xia-Ji Liu, Hui Hu, and Peter D. Drummond, Phys. Rev. Lett. \textbf{102}, 160401 (2009).
\bibitem{viri2} Xia-Ji Liu, Physics Reports, \textbf{524}, 37 (2013).
\bibitem{newviri} Y. Yan and D. Blume, Phys. Rev. Lett. \textbf{116}, 230401 (2016).
\bibitem{newb4} S. Endo and Y. Castin, Phys. Rev. A \textbf{92}, 053624 (2015).
\bibitem{tmat} R. Haussmann, W. Rantner, S. Cerrito, and W. Zwerger, Phys. Rev. A \textbf{75}, 023610 (2007).
\bibitem{ucontact1} F. Werner and Y. Castin, Phys. Rev. A \textbf{86}, 013626 (2012).
\bibitem{ucontact2} J. E. Drut, T. A. L\"{a}hde, and T. Ten, Phys. Rev. Lett. \textbf{106}, 205302 (2011).
\bibitem{ucontact3} K. Van Houcke, F. Werner, E. Kozik, N. Prokof'ev, B. Svistunov, preprint  arXiv:1303.6245.
\bibitem{ucontact4} E. D. Kuhnle, S. Hoinka, P. Dyke, H. Hu, P. Hannaford, and C. J. Vale, Phys. Rev. Lett. \textbf{106}, 170402 (2011).
\bibitem{bour} S. Bour, X. Li, D. Lee, U-G. Mei{\ss}ner, and L. Mitas, Phys. Rev. A \textbf{83}, 063619 (2011).
\bibitem{drut} J. E. Drut and A. N. Nicholson, J. Phys. G: Nucl. Part. Phys. \textbf{40}, 043101 (2013).
\bibitem{kasi1} J. Carlson, S. Y. Chang, V. R. Pandharipande, and K. E. Schmidt, Phys. Rev. Lett. \textbf{91}, 50401 (2003).
\bibitem{kasi2} G. E. Astrakharchik, J. Boronat, J. Casulleras, and S. Giorgini, Phys. Rev. Lett. \textbf{93}, 200404 (2004).
\bibitem{kasi3} J. Carlson and S. Reddy, Phys. Rev. Lett. \textbf{95}, 060401 (2005).
\bibitem{kasi4} A. J. Morris, P. L\'{\o}pez R\'{\i}os, and R. J. Needs, Phys. Rev. A \textbf{81}, 033619 (2010).
\bibitem{kasi5} M. Forbes, S. Gandolfi, and A. Gezerlis, Phys. Rev. Lett. \textbf{106}, 235303 (2011).
\bibitem{kasi6} V. K. Akkineni, D. M. Ceperley, and N. Trivedi, Phys. Rev. B \textbf{76}, 165116 (2007).
\bibitem{kasi7} O. Juillet, New J. Phys. \textbf{9}, 163 (2007).
\bibitem{kasi8} A. Perali, P. Pieri, and G. C. Strinati, Phys. Rev. Lett. \textbf{93}, 100404 (2004).
\bibitem{kasi9} R. Haussmann, W. Rantner, S. Cerrito, and W. Zwerger, Phys. Rev. A \textbf{75}, 023610 (2007).
\bibitem{kasi10} J. R. Engelbrecht, M. Randeria, and C. A. R. Sa de Melo, Phys. Rev. B \textbf{55}, 15153 (1997).
\bibitem{kasi11} T. Papenbrock, Phys. Rev. A \textbf{72}, 041603(R) (2005).
\bibitem{kasi12} B. Krippa, J. Phys. A \textbf{42}, 465002 (2009).
\bibitem{kasi13} P. Nikolic and S. Sachdev, Phys. Rev. A \textbf{75}, 033608 (2007).
\bibitem{kasi14} T. Schafer, C.-W. Kao, and S. R. Cotanch, Nucl. Phys. A \textbf{762}, 82 (2005).
\bibitem{kasi15} Y. Nishida and D. T. Son, Phys. Rev. Lett. \textbf{97}, 050403 (2006).
\bibitem{kasi16} Y. Nishida and D. T. Son, Phys. Rev. A \textbf{75}, 063617 (2007).
\bibitem{kasi17} J.-W. Chen and E. Nakano, Phys. Rev. A \textbf{75}, 043620 (2007).
\bibitem{kasi18} P. Arnold, J. E. Drut, and D. T. Son, Phys. Rev. A \textbf{75}, 043605 (2007).
\bibitem{kasi19} Y. Nishida, Phys. Rev. A \textbf{79}, 013627 (2009).
\bibitem{kasi20} G. A. Baker, Phys. Rev. C \textbf{60}, 054311 (1999).
\bibitem{kasi21} H. Heiselberg, Phys. Rev. A \textbf{63}, 043606 (2001).
\bibitem{kasi22} H. Hu, P. D. Drummond, and X. Liu, Nat. Phys. \textbf{3}, 469 (2007).
\bibitem{kasi23} J. Chen, Chin. Phys. Lett. \textbf{24}, 1825 (2007).
\bibitem{Ekasi1} M. Bartenstein, A. Altmeyer, S. Riedl, S. Jochim, C. Chin, J. H. Denschlag, and R. Grimm, Phys. Rev. Lett. \textbf{92}, 120401 (2004).
\bibitem{Ekasi2} T. Bourdel, L. Khaykovich, J. Cubizolles, J. Zhang, F. Chevy, M. Teichmann, L. Tarruell, S. J. J. M. F. Kokkelmans, and C. Salomon, Phys. Rev. Lett. \textbf{93}, 050401 (2004).
\bibitem{Ekasi3} J. Kinast, A. Turlapov, J. E. Thomas, Q. Chen, J. Stajic, and K. Levin, Science \textbf{307}, 1296 (2005).
\bibitem{Ekasi4} G. B. Partridge, W. Li, R. I. Kamar, Y. Liao, and R. G. Hulet, Science \textbf{311}, 503 (2006).
\bibitem{Ekasi5} J. T. Stewart, J. P. Gaebler, C. A. Regal, and D. S. Jin, Phys. Rev. Lett. \textbf{97}, 220406 (2006).
\bibitem{Ekasi6} J. Joseph, B. Clancy, L. Luo, J. Kinast, A. Turlapov, and J. E. Thomas, Phys. Rev. Lett. \textbf{98}, 170401 (2007).
\bibitem{Ekasi8} L. Luo and J. E. Thomas, J. Low Temp. Phys. \textbf{154}, 1 (2009).
\bibitem{carlsonNew} J. Carlson, S. Gandolfi, K. E. Schmidt, and S. Zhang, Phys. Rev. A \textbf{84}, 061602(R) (2011).
\bibitem{b4w1} D. Rakshit, K. M. Daily, and D. Blume, Phys. Rev. A \textbf{85}, 033634 (2012).
\bibitem{b4w2} V. Ngampruetikorn, M. M. Parish, and J. Levinsen, Phys. Rev. A \textbf{91}, 013606 (2015).
\bibitem{nkasi1} M. G. Endres, D. B. Kaplan, J-W. Lee, and A. N. Nicholson, Phys. Rev. A \textbf{87}, 023615 (2013).
\bibitem{nkasi2} X. Li, J. Koloren\v{c}, and L. Mitas, Phys. Rev. A \textbf{84}, 023615 (2011).
\bibitem{nkasi3} J. Carlson, S. Gandolfi, K. E. Schmidt, and S. Zhang, Phys. Rev. A \textbf{84}, 061602(R) (2011).
\bibitem{nkasi4} M. M. Forbes and R. Sharma, Phys. Rev. A \textbf{90}, 043638 (2014).
\end{references}
\end{document}